# A color-difference formula for evaluating color pairs with no separation - $\Delta E_{NS}$

FERESHTEH MIRJALILI,[1] MING RONNIER LUO,[1,2,*] GUIHUA CUI,[3] AND JAN MOROVIC[4]

[1]*State Key Laboratory of Modern Optical Instrumentation, Zhejiang University, Hangzhou 310027, China*
[2]*School of Design, University of Leeds, Leeds LS2 9JT, UK*
[3]*School of Physics & Electronics Information Engineering, Wenzhou University, Wenzhou 325035, China*
[4]*HP Inc., Barcelona, Catalonia, Spain*
*\*m.r.luo@zju.edu.cn*

**Abstract:** All color-difference formulas are developed to evaluate color differences for pairs of stimuli with hair-line separation. In printing applications, however, color differences are frequently judged between a pair of samples with no-separation because they are printed adjacent on the same piece of paper. A new formula, $\Delta E_{NS}$ has been developed for pairs of stimuli with 'no-separation' (NS). An experiment was conducted to investigate the effect of different color-difference magnitudes using sample pairs with NS. 1,012 printed pairs with NS were prepared around 11 CIE recommended color centers. The pairs, representing four color-difference magnitudes of 1, 2, 4 and 8 CIELAB units were visually evaluated by a panel of 19 observers using the gray-scale method. Comparison of the present data based on pairs with NS, and previously generated data using pairs with hair-line separation, showed a clear separation effect. A new color-difference equation for the NS viewing condition ($\Delta E_{NS}$) is proposed by modifying the CIEDE2000 formula. The separation effect can be well described by the new formula. For a sample pair with NS, when the CIEDE2000 color difference is less than 9.1, a larger color difference leads to a larger lightness difference, and thus the total color difference increases. When the CIEDE2000 color difference is greater than 9.1, the effect is opposite, i.e. the lightness difference decreases, and thus the total color difference also decreases. The new formula is recommended for future research to evaluate its performance in appropriate applications.

## 1. Introduction

Since the recommendation of CIELAB color space and the associated color-difference formula by the Commission Internationale de l'Eclairage (CIE) in 1976 [1], much effort has been devoted to try to improve its uniformity. Despite its simplicity, there is relatively poor correlation between color differences predicted by CIELAB and their equivalent visual judgments, especially for small to medium color differences, i.e. $\Delta E^*_{ab}<5$ [2].

In 1987, CIE recommended to collect color-difference data surrounding five color centers including gray, red, green, blue and yellow for coordinated research on color-difference evaluation [3]. Later, the number of color centers was extended to 19, of which the additional 12 centers provide extended coverage of the color gamut [4]. Following the CIE guidelines [3,4], several sets of visual data were collected using surface samples viewed under typical industrial viewing conditions, in order to develop new color-difference formulas and investigate the effect of various parametric factors such as the color of the background, the illumination, the magnitude of color difference, the stimulus size, the separation between the two samples, etc. A brief account of some important color-difference formulas and visual data sets is given below.

The CMC(l:c) color-difference formula was one of the earliest formulas after CIELAB. It was developed by the Colour Measurement Committee (CMC) of the Society of the Dyers and Colourists (SDC) based on experimental results obtained using textile samples [5]. In this formula, '*l*' refers to the lightness and '*c*' refers to the chroma weighting factor. Although CMC(l:c) performed better than CIELAB for small color differences, it did not perform well comparing with those developed in the later stage [6].

In an attempt to overcome the problems associated with CMC(l:c), the BFD(l:c) color-difference formula was derived by Luo and Rigg at the University of Bradford, based on over 500 pairs of wool samples and using the gray-scale method to combine the many previously published data sets [7,8]. The gray-scale method is a standard assessment method in the textile industry for assessing color fastness [9]. In this method, the color difference between a pair of stimuli is compared visually with the lightness differences of a series of gray samples. A detailed description of the gray-scale method is given in section 2.2. This method has also been widely used in the color-difference research field (see below).

The CIE recommended the CIE94 color-difference formula [10] for field trials proposed by Berns *et al.* [11] using visual assessments of 156 glossy paint sample pairs around 19 color centers, collected using the method of constant stimuli. This data set is known as the RIT-DuPont data set. Later, Kim and Nobbs [12] investigated the weighting functions in the CIELAB color-difference formula using glossy paint samples. The corresponding data set named the Leeds data set, consists of 243 and 104 pairs accumulated using the gray scale and pair comparison methods, respectively. The outcome was the Leeds Color-Difference (LCD) equation. Using a series of 418 paint samples around the five CIE color centers, Witt [13] accumulated a new data set and studied the effect of magnitude and direction of color difference on color-difference evaluation using the gray-scale method. In 2001, using a combination of the BFD, RIT-DuPont, Leeds and Witt data sets, the CIEDE2000 color difference formula was proposed by Luo *et al.* [6] and later recommended by CIE as the standard color-difference equation [14]. This equation showed a considerable improvement over previously proposed color-difference formulas.

Although the above formulas can accurately predict perceptual color differences, they do not have an associated color space, because they are all modifications of CIELAB. Moreover, they do not consider a change in the viewing parameters such as the luminance of the adaptation field, the magnitude of the color difference, the separation effect, etc. In an attempt to develop a new, perceptually uniform color space, Luo *et al.* [15] revised their color appearance model, CIECAM97s to improve its accuracy and make it more simple. They proposed a new color appearance model which was adopted by the CIE as the CIECAM02 color appearance model [16]. Luo *et al.* later derived three uniform color spaces based on CIECAM02, to predict small color differences, to predict large color differences, and a combination of both, named CAM02-SCD, CAM02-LCD, and CAM02-UCS, respectively. CIECAM02 and CAM02-UCS have been widely used in many applications. However, some mathematical problems in the chromatic adaptation transform in CIECAM02 were found. Li *et al.* [17] have recently revised the CIECAM02 model to solve these problems and have proposed a new color appearance model, CAM16, and its corresponding uniform color space, CAM16-UCS. It is to be hoped that this model will receive CIE recommendation at some time in the near future [18].

Morillas and Fairchild [19] recently proposed two color-difference metrics based on color-discrimination ellipsoids derived from the RIT-DuPont data set. They found that the performance of both metrics is significantly dependent on the magnitude of the color difference. After optimizing the equations with a power factor and a scaling factor, with respect to the magnitude of the color difference, both formulas outperformed CIEDE2000. However, the new metrics are complex and computationally expensive, with 21 parameters to be computed.

Another stream of color-difference research has been to study the influence of various parametric factors such as the separation and the color-difference magnitude, on the perceived color difference. CIE has recommended a set of 'reference' viewing conditions for assessing color differences, i.e. a pair of hair-line divided samples under a D65 simulator at 1000 lux, viewed by observers with normal color vision, object viewing mode, stimulus size of more than 4° subtended visual angle, color-difference magnitude of 0 to 5 CIELAB units and visually homogeneous sample structure [4]. Since the two physical samples need to be juxtaposed, there would inevitably be a fine dividing line, known as a hair-line, between the two samples. The psychophysical method for data acquisition was not specified by the CIE. However, most of the available data sets have been generated using the gray-scale or constant stimuli methods.

The effect of separation on perceived color difference is an important aspect of color-difference evaluation. When evaluating color differences, three types of separation can be considered: hair-line, gap and 'no-separation' (NS). A hair-line is the virtual line which appears between a pair of samples when they are placed side-by-side. On the other hand, a gap is a larger spatial distance between a pair of samples such that the background can be clearly seen between the two samples. For samples with NS, the two samples are juxtaposed in a way that observers see the change of color from one sample to the other without any interference of a hair-line or a gap. In order to achieve the pairs with NS in this work, the two samples were printed side by side on the same substrate as one physical specimen.

The conventional color-difference formulas are all developed based on pairs with hair-line separation. However, in printing applications it is usually the case that colors are printed adjacent to one another, with no discernible gap, on the same medium, e.g. paper (documents), card (packaging) or vinyl (display advertising). In this specific application, some problems with respect to the ineffectiveness of color-difference formulas have been reported and it is this condition that is investigated in this paper.

In one of the earliest studies on the effect of separation on color-difference perception of painted specimens, Witt [20] used a large gap of 3 mm width between the two samples, constituting an angular subtense of 0.5°. The results revealed that a correcting gap-factor should be considered in the color-difference formulas. Witt also demonstrated that the gap-factor decreases with increasing lightness of the samples.

Guan and Luo [21,22] carried out an extensive study on two parametric effects: gap and magnitude of color difference. They compared the effect of a hair-line and a large (3-inch) gap on the color-difference judgments of 75 wool sample pairs. The mean color difference of the whole data set was 3 CIELAB units. It was found that, although the visual differences of sample pairs with large separation were approximately 11% smaller than those for pairs having hair-line separation, the separation effect was not as obvious as that found by Witt [20]. They also investigated the parametric effect of sample separation using large color differences (a mean of 13 CIELAB units). Again, the sample pairs with a large gap showed smaller color differences than the pairs with hair-line separation. The results of comparison of the chromaticity discrimination ellipses between small and large color-differences implied that there might be a large difference between small and large color-difference perception.

In a similar study, Xin *et al.* [23] used dyed cotton sample pairs around five CIE color centers. The average color difference of the sample pairs was approximately 5 CIELAB units. They also demonstrated that the perceived color-difference of pairs with a hair-line separation was 8% larger than those of the pairs with a 3-inch separation.

Xu and Yaguchi [24] studied the effect of color-difference magnitude on inter-observer variability. Using a new visual data set ranging from small to large color differences around the five CIE color centers on a CRT display, they showed that inter-observer variability for small color differences was approximately 40% inferior to that of large ones, implying less precision in color discrimination judgment for small differences.

Using self-luminous sample pairs with no-separation, 1-pixel, 2-pixel, and large gaps on a CRT display, Cui *et al.* [25,26] found that changing the separation size between the two samples of a pair had only a small effect on the perceived color-difference, but it did change the weighting factor between the lightness difference and the chromatic difference. However, they did find a distinct difference between pairs with and without separation (to be discussed later).

The 'gap effect' has been also extensively studied by color-vision scientists. Pioneering research on the effect of a gap on color discrimination was conducted by Boynton *et al*. [27]. They defined the gap effect as a phenomenon of altered discriminability due to a separation between the fields. They referred to it as a 'positive' or a 'negative' effect depending on whether the discriminability was improved or impaired, respectively. Their results showed that by introducing the gap between two color fields in a pair, the chromatic discriminability improved, while the luminance discriminability was impaired. It was found by later research that chromatic discriminability improved when only the signal of the short-wavelength (S) cones was varying. For discrimination where only the ratio of the signals of the long-wavelength (L) and middle-wavelength (M) cones was varying, a small gap effect was observed [27,28]. Eskew [29] found that the gap effect was reduced by increasing the exposure time. This may indicate that the effect of the gap on chromatic discrimination might not be as significant as its effect on lightness discrimination. Note that in most of these studies, it was difficult to achieve a no-separation arrangement because some kind of a dividing line could still be observed between the two stimuli.

With the above in mind, two goals were set for this research: to investigate the difference between pairs of samples with NS and pairs of samples with hair-line separation, and to study the effect of color-difference magnitude on color-difference evaluation.

## 2. Experimental methods

### 2.1 Sample preparation

Eleven CIE color centers, distributed uniformly in CIELAB color space, were chosen for this study. These color centers were gray, red, high-chroma orange, yellow, high-chroma yellow-green, green, high-chroma green, blue-green, blue, high-chroma purple, and black. Table 1 gives the CIELAB values of the chosen color centers.

**Table 1. CIELAB color attributes $L^*$, $a^*$, $b^*$ of the 11 color centers calculated using the CIE illuminant D65/1964 colorimetric observer combination**

|    | Color center             | $L^*$ | $a^*$ | $b^*$ | $C_{ab}^*$ | $h_{ab}$ |
|----|--------------------------|-------|-------|-------|------------|----------|
| 1  | Gray                     | 61.1  | -3.2  | 3.2   | 4.5        | 135      |
| 2  | Red                      | 41.0  | 33.2  | 25.5  | 41.9       | 38       |
| 3  | High-chroma orange       | 60.3  | 33.0  | 64.3  | 72.2       | 63       |
| 4  | Yellow                   | 84.1  | -6.7  | 50.4  | 50.9       | 98       |
| 5  | High-chroma yellow green | 63.2  | -29.3 | 44.1  | 53.0       | 124      |
| 6  | Green                    | 56.2  | -32.5 | 4.9   | 32.8       | 172      |
| 7  | High-chroma green        | 56.0  | -45.7 | 5.7   | 46.1       | 173      |
| 8  | Blue green               | 50.6  | -18.7 | -6.9  | 19.9       | 200      |
| 9  | Blue                     | 37.0  | -1.3  | -27.9 | 28.0       | 267      |
| 10 | High-chroma purple       | 45.4  | 18.9  | -25.0 | 31.4       | 307      |
| 11 | Black                    | 29.8  | -3.1  | 2.3   | 3.8        | 143      |

For each color center, a systematic distribution of samples around the center in CIELAB color space was designed and produced. A group of 7, 7 and 9 pairs were prepared in $L^*a^*$, $L^*b^*$ and $a^*b^*$ planes for each color center, respectively. In each plane, the two color attributes varied while the third one was approximately constant, i.e. $\Delta b^*$, $\Delta a^*$ or $\Delta L^*$ in $L^*a^*$, $L^*b^*$ or $a^*b^*$ planes, respectively, was always approximately zero. Note that $\Delta$ designates the difference between the sample and the color center. Additionally, the pairs had four levels of color-difference magnitudes, namely 1, 2, 4 and 8 CIELAB units. These levels are denoted as $\Delta E_M$ =1, 2, 4 and 8 respectively. In total, 1,012 pairs of samples were prepared for each color center.

It can be seen from Table 1 that the color centers cover a wide range of $L^*$ [30, 84], $a^*$ [-46, 33] and $b^*$ [-27, 64]. Figures 1(a)-(c) illustrate the distribution of the color centers in CIELAB $a^*b^*$, $L^*a^*$ and $L^*b^*$ planes, respectively. Figure 1(d) shows the sample distribution around the gray center for $\Delta E_M$ of 8 CIELAB units. $\Delta L^*$, $\Delta a^*$ and $\Delta b^*$ are the differences between the color center and the sample.

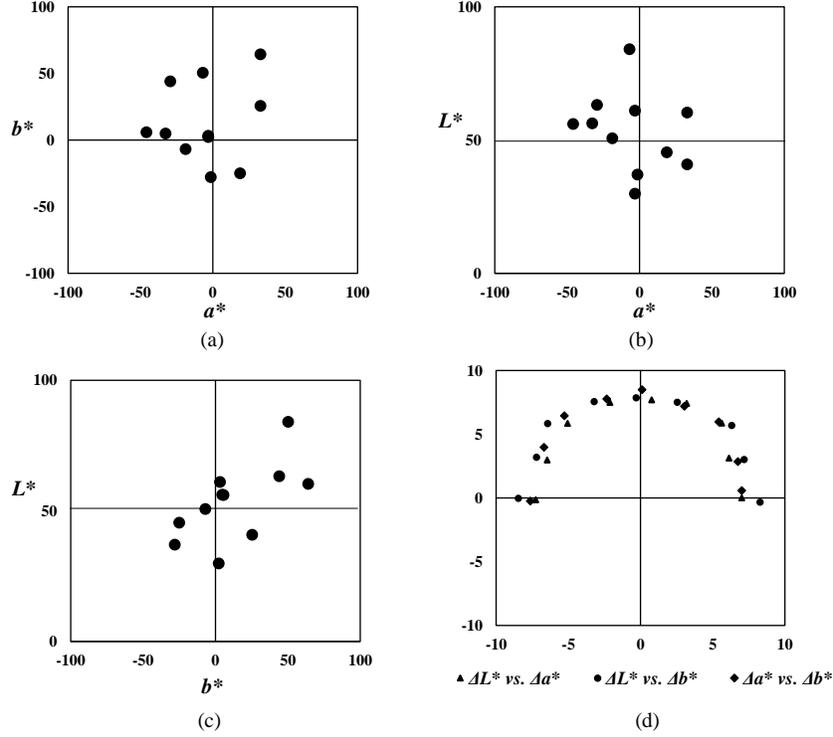

Fig. 1. Distribution of the 11 CIE color centers in CIELAB (a) $a^*b^*$, (b) $L^*a^*$, and (c) $L^*b^*$ planes; (d) the distribution of the samples around the gray center in $L^*a^*$, $L^*b^*$ and $a^*b^*$ planes for a color-difference magnitude of 8 CIELAB units.

The sample pairs were printed on an HP Latex 365 Printer (HP, Barcelona, Spain) on an Avery Dennison matt white polymeric Self-Adhesive Vinyl substrate with CMYKcm inks. For each pair, the color center and its corresponding sample were printed adjacent to each other on the same substrate such that there was no separation between them. Each pair had a size of 8 cm x 8 cm, i.e. each sample in the pair had a size of 4 cm x 8 cm, and an approximate vertical field of view of 4°. The spectral reflectance of the samples was measured using an X-Rite SpectroEye spectrophotometer (X-Rite, Grand Rapids, US). This portable instrument has 45°:0° measuring geometry and measures the spectra in the range of 380 nm – 730 nm with a spectral resolution of 10 nm. The short-term repeatability of the spectrophotometer and the uniformity of the printed samples were assessed before sample measurements. The former was evaluated by measuring a high-chroma green sample continuously 40 times within approximately 5 minutes. The latter was evaluated by measuring five points on the sample. For both tests, the mean color difference from the mean (MCDM) metric, in CIELAB units, was used [30]. The short-term repeatability and the sample uniformity were 0.0006 and 0.04 MCDM, respectively. These values indicate that the instrument has a high repeatability performance and the samples have good uniformity. In addition, for each sample pair, the sample representing the color center was measured only at one point, in the center, while the difference sample was measured at two points and the average of the two measurements calculated. The measurements were repeated during and at the end of the experiments to make sure that the samples underwent no color fading.

## 2.2 Visual assessment of color difference

The widely used gray-scale method was used for the visual assessment of the color difference [9,31]. The gray scale was prepared using the same material as the samples. The color specification of the patches is given in Table 2. The gray scale consisted of 9 samples, each with a different lightness level including a 'standard' sample (i.e. the darkest sample or sample 1) and 8 gray-scale samples. The gray-scale samples were prepared in such a way that the differences between the 'standard' and each of the samples (samples 1 to 8) were essentially only a lightness difference: i.e. no variation in chroma $C_{ab}*$ or hue angle $h_{ab}$. Eq. (1) was fitted to the data to define a relationship between the gray-scale numbers ($GS$) and their corresponding CIELAB color differences ($\Delta E_{ab}*$). This equation is commonly used in conjunction with the gray-scale method.

$$\Delta V = 0.0534 exp(0.701(GS)) \tag{1}$$

The visual color-differences reported by the observers ($\Delta V$) and the respective $\Delta E_{ab}*$ values should both increase monotonically. Therefore, Eq. (1) can be used to convert the visual differences to the corresponding $\Delta E_{ab}*$ values.

Table 2. CIELAB values of the gray-scale samples calculated using the CIE illuminant D65/1964 colorimetric observer combination

| Gray-scale number ($GS$) | $L*$ | $a*$ | $b*$ | $\Delta L*$ | $\Delta E_{ab}*$ |
|---|---|---|---|---|---|
| 1 | 41.50 | 0.05 | 1.70 | 0.00 | 0.00 |
| 2 | 41.20 | 0.12 | 1.49 | 0.29 | 0.38 |
| 3 | 41.09 | -0.01 | 1.34 | 0.40 | 0.55 |
| 4 | 42.17 | 0.13 | 1.42 | 0.68 | 0.75 |
| 5 | 43.07 | 0.09 | 1.76 | 1.58 | 1.58 |
| 6 | 45.14 | 0.09 | 1.67 | 3.65 | 3.65 |
| 7 | 48.81 | 0.04 | 1.39 | 7.32 | 7.32 |
| 8 | 56.00 | -0.50 | 1.17 | 14.51 | 14.53 |
| Standard | 41.49 | 0.04 | 1.71 | | |

A panel of 19 observers, including 10 males and 9 females, who were undergraduate and graduate students of the Zhejiang University, participated in the visual assessment experiments. They were aged from 22 to 33 years old (i.e. average age of 27.5 years with a standard deviation of 5.5) and all had normal color vision according to the Ishihara Color Vision test. The visual assessments were conducted inside a viewing cabinet equipped with a spectrum-tunable LED lighting system (Thouslite, Changzhou, China) set to simulate CIE D65 illumination. The interior of the viewing cabinet was painted Munsell N7 natural gray. The colorimetric characteristics of the D65 simulator, including the spectral power distribution (SPD), the correlated color temperature (CCT), the color rendering index (CRI) and the illuminance were measured using a JETI Specbos 1211 spectroradiometer (Jena, Germany). The light source had a CCT of 6460, a CRI of 97 and an illuminance of 960 lx. It produced no energy in the UV region.

In the psychophysical experiment, the observers sat on a chair in front of the viewing cabinet at a distance of approximately 45 cm from the samples. The illumination: viewing geometry was always approximately 0°:45°. The height of the chair was always adjusted to maintain the viewing distance and hence the viewing angle. The observers were asked to adapt to the mid-gray interior of the cabinet for 3 minutes. After adaptation, they were provided with the gray-scale samples (GS-1 to GS-8) and a 'test pair' with NS for which the color difference was to be evaluated. Figure 2 shows the gray scale samples and the test pair inside the viewing cabinet. In order to visually assess the color difference of the test pair, the observers were asked to choose one of the gray scale samples and place it next to the 'standard' having the identical color as GS-1. They had to compare the color difference of the test pair with the color difference formed by the 'standard' and the gray-scale sample, and repeat this comparison until they find the gray scale sample having the closest color difference to that of the test pair, and report its number (1, 2, …, 8). Note that the gray-scale pair was viewed under a hair-line condition throughout the experiment. Similar conditions were used to collect the visual data used in the development of CIEDE2000.

In order to evaluate the intra-observer variability, the observers assessed the color differences of the 92 pairs associated to the gray color center twice. In total, 1,012 sample pairs were visually assessed by each observer in 15 to 17 separate sessions. Each session was completed without any time restrictions, although it usually lasted 45 minutes to one hour for each observer. Overall, 20,976 visual assessments were conducted in about 300 sessions by 19 observers in 576 hours (i.e. 30 hours for each observer) within a time frame of two months.

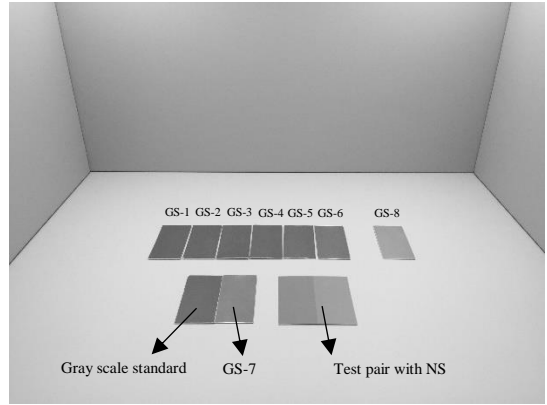

Fig. 2. Configuration of the test pair and the gray-scale samples in the viewing cabinet. [Note that the gray-scale pair at the bottom left has hair-line separation and the test pair at the bottom right has no-separation].

## 3. Results and discussion

### 3.1 Observer accuracy

The gray scale numbers reported by the observers ($GS$) were converted to the corresponding color differences ($\Delta V$) using Eq. (1). The visual color differences ($\Delta V$) were then used to evaluate the observers' accuracy and test the performance of various color-difference formulas. In order to evaluate the extent of observers' accuracy in terms of intra-observer and inter-observer variability, the standardized residual sum of squares ($STRESS$) metric [32] was used. $STRESS$ is the most widely used assessment metric used in color-difference research. By comparing two data sets $A$ and $B$, $STRESS$ can be calculated using Eq. (2):

$$STRESS = \left( \frac{\sum_{i=1}^{n}(A_i - FB_i)^2}{\sum_{i=1}^{n} F^2 B_i^2} \right)^{1/2} \times 100 \qquad (2)$$

with $F = \sum_{i=1}^{n} A_i^2 / \sum_{i=1}^{n} A_i B_i$

where *n* is the number of sample pairs and *F* is a scaling factor to adjust *A* and *B* data sets on to the same scale. The percent *STRESS* values are always between 0 and 100. Values of *STRESS* near to zero indicate better agreement between two sets of data. In color-difference studies, a *STRESS* value exceeding 35 is typically an indicator of the poor performance of the color-difference formula [33].

For intra-observer variability, the average *STRESS* value of the 19 observers was 17. This value indicates that all observers were reasonably internally consistent. Table 3 presents the inter-observer variability of the observers in terms of *STRESS* for different color centers and color-difference magnitudes. The inter-observer variability of the 19 observers for all color centers ranged from 16 to 39 *STRESS* units with a mean value of 28 units, which is larger than the average intra-observer variability (i.e. 17 *STRESS* units) as might be expected. This value indicates a reasonable degree of consistency between the observers. The typical inter-observer variability for color-difference evaluation is around 35 *STRESS* units which has been reported by other researchers [33]. The consistent results presented here may be attributed to the characteristics of the sample set with no separation.

**Table 3. Inter-observer variability of observers in terms of *STRESS* for different color centers and color difference magnitudes**

| Color center | L*a* | L*b* | a*b* | $\Delta E_M$ | | | | Overall STRESS |
|---|---|---|---|---|---|---|---|---|
| | | | | 1 | 2 | 4 | 8 | |
| Gray | 17 | 19 | 24 | 29 | 22 | 20 | 19 | 22 |
| Red | 24 | 24 | 33 | 35 | 30 | 30 | 27 | 30 |
| High-chroma orange | 23 | 23 | 34 | 33 | 30 | 27 | 26 | 29 |
| Yellow | 23 | 26 | 39 | 44 | 35 | 27 | 24 | 29 |
| High-chroma yellow green | 25 | 25 | 36 | 41 | 31 | 28 | 24 | 28 |
| Green | 25 | 25 | 37 | 39 | 32 | 29 | 24 | 29 |
| High-chroma green | 19 | 22 | 30 | 30 | 26 | 26 | 21 | 25 |
| Blue green | 28 | 29 | 36 | 46 | 33 | 29 | 27 | 31 |
| Blue | 32 | 30 | 35 | 56 | 39 | 28 | 27 | 30 |
| High-chroma purple | 22 | 25 | 38 | 35 | 29 | 26 | 25 | 28 |
| Black | 24 | 24 | 33 | 36 | 29 | 27 | 24 | 27 |
| Mean | 24 | 25 | 34 | 39 | 30 | 27 | 24 | 28 |

Comparing the *STRESS* values for different color centers in Table 3 shows that the lowest average *STRESS* value belongs to the gray center, suggesting that the assessment of the color difference of the gray stimuli might be easier for observers than the other color centers. The results shown in Table 3 also indicate that observers showed better performance in the assessment of the color difference of sample pairs in the *L*a** and *L*b** planes as compared to the *a*b** plane, given the average *STRESS* values of 24, 25 and 34, respectively. The mean *STRESS* value of the observations decreases with increasing color-difference magnitude. In other words, a higher observation variability is found when assessing the color difference of pairs having smaller color-differences.

*3.2 Correlation between various visual data sets*

The results of experiments on color difference can be conveniently summarized and compared as chromaticity discrimination ellipses. For each color center, color discrimination can be represented by the ellipsoid equation:

$$\Delta E^2 = k_1 \Delta a^{*2} + k_2 \Delta a^* \Delta b^* + k_3 \Delta b^{*2} + k_4 \Delta a^* \Delta L^* + k_5 \Delta b^* \Delta L^* + k_6 \Delta L^{*2} \qquad (3)$$

where coefficients $k_1$ to $k_6$ are optimized to give the lowest *STRESS* value between the color differences calculated using Eq. (3) and the visual data ($\Delta V$) for each color center. Setting $\Delta E$ to unity allows calculation of $\Delta L^*$, $\Delta a^*$ and $\Delta b^*$ values of an ellipsoid corresponding to $\Delta V$ of

1. Setting $\Delta L^*$ to zero allows the corresponding ellipse in $\Delta a^*\Delta b^*$ plane to be calculated. The ellipse equation for the present data set, referred to as the ZJU (Zhejiang University) data set, was fitted for each color-difference magnitude (i.e. $\Delta E_M$ of 1, 2, 4 and 8), separately.

Figure 3 plots the fitted ellipses in the $a^*b^*$ plane. There exists a general agreement between various color-difference magnitudes in terms of the ellipse shape and orientation. However, the larger ellipses are obtained for larger color-differences. Moreover, the smallest ellipses are located at the origin and the ellipse size increases by increasing the chroma. Also, most of the ellipses are oriented towards the origin except for those in the blue region.

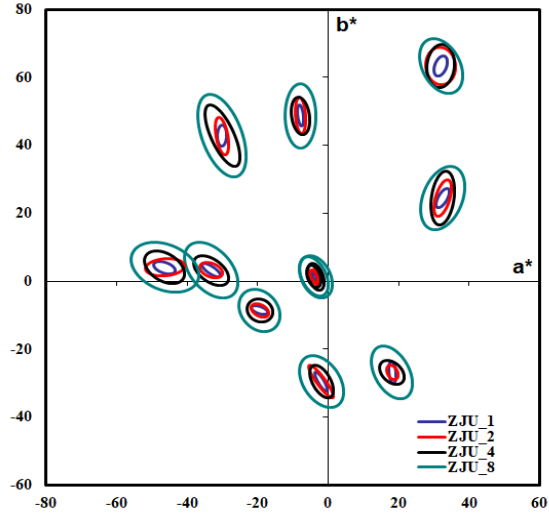

Fig. 3. Chromaticity ellipses of 11 color centers for various color difference magnitudes in $a^*b^*$ plane; blue: $\Delta E_M = 1$, red: $\Delta E_M = 2$, black: $\Delta E_M = 4$, green: $\Delta E_M = 8$.

The ZJU ellipses were compared with four previously published data sets (Witt, RIT–DuPont, Cheung and Rigg [34], and Cui *et al.*). It is reasonable to compare these data sets because they were all generated based on the five CIE color centers.

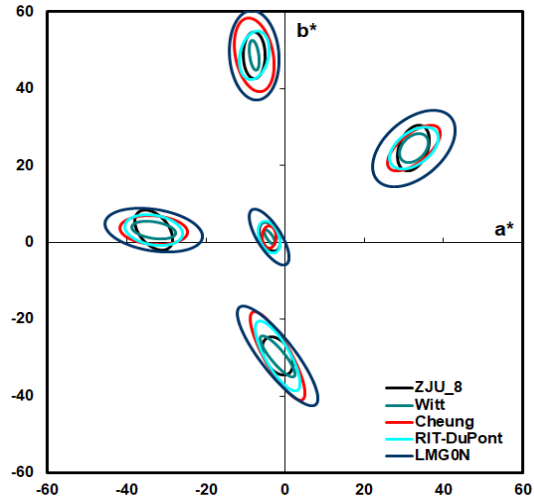

Fig. 4. Chromaticity ellipses of the five CIE color centers for different data sets plotted in an $a^*b^*$ plane; black: ZJU_8, green: Witt, red: Cheung and Rigg, cyan: RIT-DuPont, blue: Cui-LMG0N.

Figure 4 shows the ellipses fitted to the visual differences obtained for the color-difference magnitude of 8, i.e. ZJU_8, together with the ellipses of the above mentioned four data sets, in an $a^*b^*$ diagram. The ZJU_8 ellipses were chosen because they always gave the best agreement with the ellipses from the other studies. The LMG0N subset representing the pairs with NS was selected from Cui's data set for comparison with the present data. The $F$ parameter in Eq. (2) was used as the scaling factor for adjusting the sizes of the ellipses from different data sets. It automatically adjusts the $\Delta E$ in Eq. (3) to have the same size as the visual data.

Most of the data sets were generated using surface colors, except for Cui *et al.* data which was produced based on CRT colors, including 16 subsets, varying in sample size, background color, separation and color of separation, amongst which the LMG0N and the LMG1B subsets are of interest in this work. The LMG0N subset was generated against a mid-gray background with no separation. The LMG1B subset, however, was generated using the same conditions as LMG0N except for a 1-pixel black dividing line between the samples. (Note that the other subsets were not used because they were generated using different colored backgrounds or larger separation distances.) Figure 5 shows the correlation between the ZJU data set and the other data sets in terms of the *STRESS* parameter.

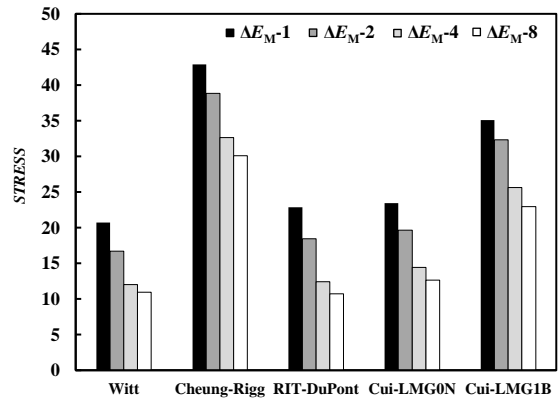

Fig. 5. The correlation between the ZJU data set and the Witt, Cheung and Rigg, RIT-DuPont and Cui *et al.* data sets in terms of *STRESS*.

Although the ZJU ellipses agree reasonably well with the other data sets, it can be seen in figure 5 that the ZJU_8 subset agrees the best with the other sets in terms of the shape and orientation of the ellipses, but not the size. This is most likely due to the smaller inter-observer variability associated to this subset (see Table 3). Additionally, irrespective of the magnitude of color difference, the ZJU ellipses agree the best with the Witt and RIT-DuPont ellipses, and the worst with Cheung and Rigg's ellipses. This discrepancy is most likely attributed to the texture of the surface colors used in these studies. Although the pairs in the three data sets were assessed under the hair-line condition, due to having unsmooth edges, the textile pairs in the Cheung and Rigg's work presented a wider gap than the paint samples in the RIT-DuPont and the Witt sets. This suggests that even the width of the hair-line has a big impact on the visual results.

Comparing with the LMG0N and the LMG1B subsets from Cui's data set, the ZJU data set agrees markedly better with LGM0N, which was produced using the sample pairs with NS. This suggests that separation has a great impact on color-difference perception and there is not much difference between color-difference perception of surface and self-luminous colors with NS.

*3.3 Performances of various color difference formulas*

Eq. (4) shows a generic color-difference formula, including the lightness, chroma and hue parametric factors, $k_L$, $k_C$ and $k_H$ for the lightness, chroma and hue, respectively, designed to consider different viewing parameters such as texture, background, separation, etc.

$$\Delta E = \sqrt{\left(\frac{\Delta L}{k_L S_L}\right)^2 + \left(\frac{\Delta C}{k_C S_C}\right)^2 + \left(\frac{\Delta H}{k_H S_H}\right)^2 + R_T \left(\frac{\Delta C}{k_C S_C}\right)\left(\frac{\Delta H}{k_H S_H}\right)} \qquad (4)$$

where $\Delta L$, $\Delta C$ and $\Delta H$ are differences in lightness, chroma and hue, $k_L$, $k_C$, and $k_H$ are the parametric factors and the $S_L$, $S_C$, and $S_H$ are the weighting functions for the lightness, chroma, and hue components, respectively. $R_T$ is the rotation function provided to improve the performance of the color-difference formula when fitting chromatic differences in the blue region of color space [6]. Note that the parametric factors were designed for different applications, e.g. for the CIEDE2000 and CIE94 formulas, $k_L=k_C=k_H=1$ for samples having a smooth surface such as paint and plastic patches, while $k_L=2$ and $k_C=k_H=1$ for rough surfaces such as textile samples.

The performance of a set of color-difference formulas including CIELAB, CMC, CIEDE2000, CIE94, CAM02-UCS and CAM16-UCS was tested using the present data set. In order to investigate the separation and color-difference magnitude effects on the performance of each formula, three forms of color-difference formula were tested: the original, the power-corrected [35], and the parametric factor-optimized [21] equations. It is expected that the last two modifications should enhance the performance of all formulas.

In the first test, the original form of each color-difference formula in which $k_L=k_C=k_H=1$ was used. The performance of each formula in predicting the visual differences was then evaluated in terms of the *STRESS* parameter. The respective *STRESS* values are compared as bar charts in Figure 6(a) and it can be seen that in original form when $k_L=k_C=k_H=1$, all formulas markedly outperformed the CIELAB and the CMC formula. CIE94 performed the best overall followed by CAM16-UCS, CAM02-UCS and CIEDE2000.

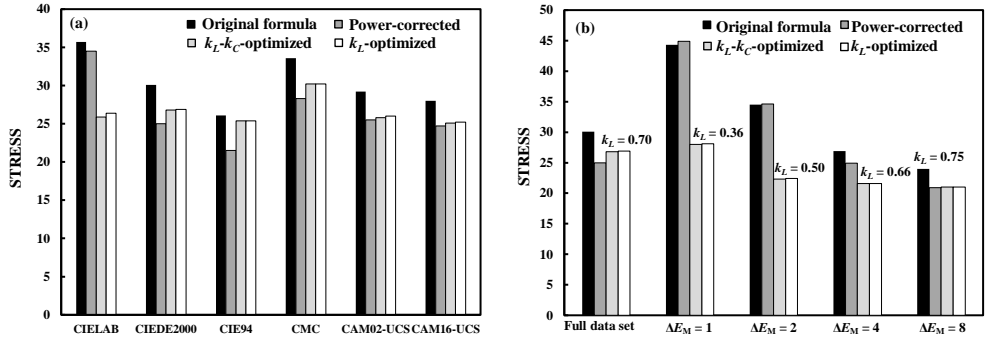

Fig. 6. The performance of the original, the power-corrected and the parametric factor-optimized color-difference formulas in terms of *STRESS* for (a) six formulas; (b) CIEDE2000.

One of the advantages of using the *STRESS* parameter to evaluate the strength of the relationship between the perceived and predicted color differences is the possibility of implementing the *F*-test using the *STRESS* values, to test the statistical difference between two color-difference formulas [32]. For two given color-difference formulas $DE_1$ and $DE_2$, the *F* value can be calculated by Eq. (5):

$$F = \frac{STRESS^2_{DE_1}}{STRESS^2_{DE_2}} \qquad (5)$$

The *F* value was used to compare the performace of the formulas after each modification. For the present data set, the critical *F* value, $F_C$, for the two-tailed *F* distribution with a 95% confidence level and (∞,∞) degrees of freedom is 1. Considering that the number of the samples was large (N=1,012), an infinite number of degrees of freedom could be assumed. No significant difference was found between the CAM02-UCS, the CAM16-UCS and the CIEDE2000 formulas according to the *F*-test.

The next test was to apply a power factor to the formulas. Huang *et al.* [35] found that introducing a single power factor could lead to an overall improvement in the performance of the formula regardless of the color-difference magnitude. As was expected, the performance of all formulas improved slightly after power-correction. However, again according to the *F*-test, there was no significant difference between the three improved formulas.

In the last test, the color-difference formulas were modified by optimizing the chroma and lightness parametric factors, $k_C$ and $k_L$, with $k_H$ =1, to give the best fit to the visual differences. Again, the performance of all formulas improved according to the *F*-test. However, it was found that the chroma parametric factor $k_C$ is always larger than $k_L$ indicating that all formulas predicted a larger lightness difference compared to the chroma difference with the hue difference in between. For all formulas except CIELAB, the $k_C$ values were close to 1, ranging from 0.82 to 0.93, while the $k_L$ values were always less than 1. This means that the chroma and hue differences were well balanced (i.e. $k_C \approx k_H$ =1), and only the lightness difference affected the total color-difference. A $k_L$ value less than one results from a larger perceived lightness difference; hence a larger total color-difference is perceived for pairs with NS (see Eq. 4). Again, this behavior might be attributed to the separation effect, i.e. the larger perceived color-difference which is mainly a lightness difference when there is no hair-line or separation between the samples.

In order to test this premise, all formulas were modified only for the $k_L$ factor with $k_C= k_H =1$. Figure 6(a) shows that the performance of all modified formulas improved. Although the improvement was not significant according to the *F*-test, this result still indicates that only applying the $k_L$ factor should be sufficient to describe the effect of color-difference magnitude. Figure 6(b) illustrates the performance of the CIEDE2000 color-difference formula for the ZJU data set and various color difference magnitudes together with the corresponding optimized $k_L$ values. It can be seen that all the optimized $k_L$ values are less than one and proportional to the size of the color difference. Note that CIEDE2000 with $k_L= k_C=k_H =1$ was developed using all the previous data which were generated under the hair-line viewing condition. For the current results, however, the $k_L$ values less than one indicate that there is a parametric effect due to the separation.

The present results demonstrated that the perceived lightness difference of pairs with NS is larger than the perceived lightness difference of pairs with hair-line separation. This result is in line with the findings of Boynton and his colleagues [27]. However, we found that chromaticity (i.e. hue and chroma) difference appears to be less affected by separation.

*3.4 Developing a color difference equation for pairs with NS*

As demonstrated in the previous section, changing the size of the color difference affects the total color-difference perception when there is no hair-line or gap between the samples. The perceived color-difference for pairs with NS is larger and it is mainly due to the lightness difference while the hue and chroma differences are well balanced. Considering this effect, a new equation for the lightness difference parametric factor is proposed as a linear function of color difference (*ΔE*). The new lightness difference parametric factor ($D_L$) is given in Eq. (6):

$$D_L = a\Delta E + b \qquad (6)$$

where *ΔE* is the color difference, and *a* and *b* are constants to be optimized. The optimized *a* and *b* values for each tested color difference formula are given in Table 4. Note that $D_L$ values are less than 1.0 unless the color difference exceeds 15.6, 9.1, 8.3, 10.3 and 10.3 for CIELAB,

CIEDE2000, CIE94, CAM02-UCS and CAM16-UCS, respectively. In other words, the lightness difference is perceived to be more visible than the chroma and hue differences until reaching to a certain level of color difference, after which the perceived lightness difference starts to be less important.

Table 4. Optimized *a*, *b*, *c* and *d* coefficients for various color difference formulas

| Color difference formula | *a* | *b* | *c* | *d* |
|---|---|---|---|---|
| CIELAB | 0.05 | 0.22 | 0.72 | 0.95 |
| CIEDE2000 | 0.08 | 0.27 | 0.70 | 0.91 |
| CIE94 | 0.08 | 0.34 | 0.73 | 0.94 |
| CAM02-UCS | 0.07 | 0.28 | 0.72 | 0.93 |
| CAM16-UCS | 0.07 | 0.27 | 0.73 | 0.93 |

To find the formula with the highest performance, three modified versions were proposed for each color difference formula: the magnitude-corrected equation, $\Delta E_1$, the power-corrected equation, $\Delta E_2$, and the magnitude-power-corrected equation, $\Delta E_3$. Their generic forms are given in Eqs. (7) to (9).

$$\Delta E_1 = \sqrt{\left(\frac{\Delta L}{D_L}\right)^2 + (\Delta C)^2 + (\Delta H)^2 + RT} \tag{7}$$

$$\Delta E_2 = \left[\sqrt{(\Delta L)^2 + (\Delta C)^2 + (\Delta H)^2 + RT}\right]^c \tag{8}$$

$$\Delta E_3 = \left[\sqrt{\left(\frac{\Delta L}{D_L}\right)^2 + (\Delta C)^2 + (\Delta H)^2 + RT}\right]^d \tag{9}$$

where *RT* is the rotation function which need to be set to zero for all formulas except for CIEDE2000, and *c* and *d* are power factors which were obtained by minimizing the *STRESS* values between each formula and the present visual data. The optimized values are shown in Table 4. Comparing the optimized coefficients in Table 4 indicates that the coefficients do not vary much across the color-difference formulas. The performance of the five formulas after optimization was tested in terms of *STRESS* and the results are compared in Figure 7(a). Additionally, the performance of the modified CIEDE2000 formula is summarized in Figure 7(b).

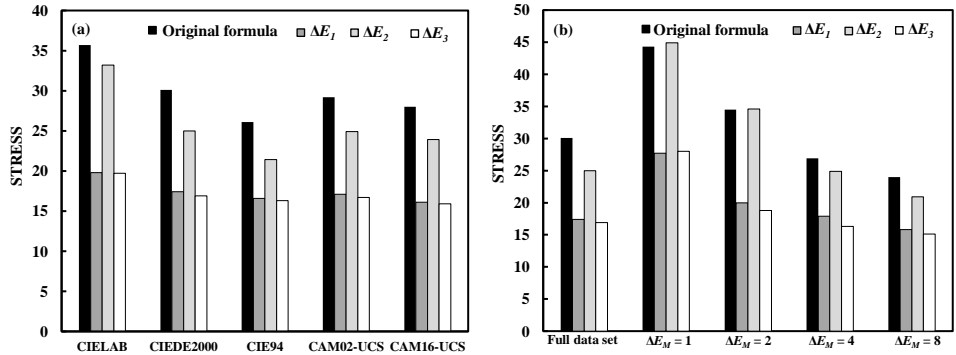

Fig. 7. Performance of original and modified color difference formula in terms of *STRESS* for (a) five formulas; (b) CIEDE2000.

Both power correction and magnitude correction are intended to improve the performance of color-difference equations. By comparing the *STRESS* values in Figure 7, it can be seen that although power correction enhanced the performance of all formulas ($\Delta E_2$), the improvement was not very obvious. The *F*-test also showed no significant difference between the original and the power-corrected ($\Delta E_2$) formulas. On the other hand, after applying the new lightness-difference parametric factor ($D_L$) in the original formula, $\Delta E_1$ showed a markedly better performance, as the *STRESS* values drastically decreased.

Further improvement in the performance of the formulas was not achieved after power-correction of $\Delta E_1$. As can be seen in Figure 7, there is not a large improvement from $\Delta E_1$ to $\Delta E_3$ as the corresponding *STRESS* values are very close. Again, CIE94 followed by CAM16-UCS and CIEDE2000 performed very well and all formulas outperformed CIELAB.

It is encouraging that the magnitude-corrected CIEDE2000 formula gave one of the most accurate predictions of all the color-difference formulas. Hence, this equation is designated as the "color-difference formula for 'no-separation' viewing condition", $\Delta E_{NS}$, which is given in Eq. (10):

$$\Delta E_{NS} = \sqrt{\left(\frac{\Delta L'}{D_L}\right)^2 + (\Delta C')^2 + (\Delta H')^2 + R_T(\Delta C')(\Delta H')} \qquad (10)$$

with $D_L = 0.08\Delta E_{00} + 0.27$,

where $\Delta L'$, $\Delta C'$ and $\Delta H'$ are the CIEDE2000 terms for lightness, chroma and hue differences, $R_T(\Delta C')(\Delta H')$ is the interactive term between chroma and hue differences, and $\Delta E_{00}$ is the CIEDE2000 color difference. These terms are calculated according to the same procedure used to calculate the CIEDE2000. A worked example of the calculation of $\Delta E_{NS}$ is given in the Appendix.

Eq. (10) can well describe the visual phenomena observed in the experiment. When $\Delta E_{00}$ is smaller than 9.1, a larger color-difference magnitude results in a higher value of $D_L$, leading to a lower lightness difference ($\Delta L'/D_L$), and a lower $\Delta E_{NS}$ value. This implies that the border between the two samples is important for judging the color difference. A clear perceived border will reduce the perceived color difference. This is in agreement with the findings of Cui *et al.* [25]. For $\Delta E_{00}$ values larger than 9.1, the effect is opposite, i.e. a larger color difference will lead to a larger $D_L$, but results in a smaller ($\Delta L'/D_L$) and hence a smaller $\Delta E_{NS}$. However, further experiment is required to verify the latter conclusion.

$\Delta E_{NS}$ is proposed for applications where there is no hair-line or separation gap between the sample pairs under judgment. The formula is now being extensively tested by the HP Inc. and the results of its performance evaluation will be reported in the near future.

## 4. Conclusions

Using a series of printed color-difference pairs without separation, a comprehensive color discrimination data set was accumulated and the effect of separation and color-difference magnitude on the performance of various color-difference formulas investigated. Modifying some of the advanced color-difference formulas by optimizing the lightness parametric factor, $k_L$, resulted in an improvement in the performance of the formulas. The findings imply that for pairs with NS, the lightness difference has the major contribution to the total color difference, although such an effect is reduced by increasing the size of the color difference. Based on the results, a new lightness-difference parametric equation is proposed as a linear function of the color difference. The new function has been applied to various color-difference formulas and the performance of all tested formulas was markedly improved. A new color-difference formula based on CIEDE2000 was developed for sample pairs with NS, covering a wide range of color-difference magnitudes smaller than 9.1 CIEDE2000 units. The new equation is designated as the color-difference formula for 'no-separation' viewing condition: $\Delta E_{NS}$. Further research should be carried out to verify the present results, especially for color differences larger than 9.1 $\Delta E_{00}$ units.

*Acknowledgments*

We would like to thank the National Science Foundation of China (Project No. 61775190), and the HP Inc. (Barcelona, Spain) for their support. We also thank all the observers who patiently participated in the visual assessment experiments.

## APPENDIX    $\Delta E_{NS}$ worked example

This example shows how to calculate the color difference between a standard (S) and a sample patch (P) using the $\Delta E_{NS}$ equation. Two sets of input data are given. The *XYZ* tristimulus values were calculated using the CIE D65 illuminant and the 1964 standard colorimetric observer.

**Table A1: The input values for the worked examples**

| Reference White | | | | Pair 1 | | | | Pair 2 | | |
|---|---|---|---|---|---|---|---|---|---|---|
| $X_n$ | $Y_n$ | $Z_n$ | | $X$ | $Y$ | $Z$ | | $X$ | $Y$ | $Z$ |
| 95.78 | 100.00 | 104.61 | $S_1$ | 8.90 | 9.53 | 23.10 | $S_2$ | 58.26 | 64.26 | 24.83 |
|  |  |  | $P_1$ | 9.21 | 9.72 | 23.38 | $P_2$ | 59.10 | 64.76 | 25.50 |

**Table A2: The intermediate and final output values for the worked examples**

|  | $L^*$ | $a^*$ | $b^*$ | $C^*$ | $\Delta E_{00}$ | $D_L$ | $\Delta E_{NS}$ |
|---|---|---|---|---|---|---|---|
| $S_1$ | 36.99 | -1.92 | -29.53 | 29.59 | 0.95 | 0.35 | 1.25 |
| $P_1$ | 37.34 | -0.82 | -29.42 | 29.43 |  |  |  |
| $S_2$ | 84.1 | -7.82 | 48.76 | 49.38 | 0.61 | 0.32 | 0.80 |
| $P_2$ | 84.36 | -6.91 | 48.10 | 48.59 |  |  |  |